\documentclass[article,useAMS,usenatbib]{mn2e}

\usepackage{amsmath}
\usepackage{amssymb}
\usepackage{graphicx} 
\usepackage{sansmath}
\usepackage{flafter}
\usepackage{float}
\usepackage{fixltx2e}

\newcommand{\etamin}{\eta_\mathrm{min}}
\newcommand{\taum}{\tau_\mathrm{m}}

\newcommand{\ddp}[2]{\frac{\upartial #1}{\upartial #2}}
\newcommand{\dddp}[2]{\dfrac{\upartial #1}{\upartial #2}}
\newcommand{\lddp}[2]{\upartial #1/\upartial #2}

\newcommand{\bn}{\mathsfbf n}
\newcommand{\bom}{\boldsymbol{\omega}}
\newcommand{\eexp}{\mathrm{e}}
\newcommand{\bJ}{\mathsfbf J}
\newcommand{\bth}{\boldsymbol{\theta}}

\newcommand{\bN}{\mathsfbf N}

\newcommand{\Rb}{R_{\mathrm{b}}}

\newcommand{\Rs}{R_{\mathrm{s}}}

\newcommand{\Oms}{\Omega_\mathrm{s}}
\newcommand{\Omp}{\Omega_\mathrm{p}}
\newcommand{\Omb}{\Omega_\mathrm{b}}
\newcommand{\Omsp}{\Omega_\mathrm{sp}}
\newcommand{\omp}{\omega_\mathrm{p}}

\newcommand{\ROLR}{R_{\mathrm{OLR}}}
\newcommand{\ROUH}{R_{\mathrm{OUH}}}

\newcommand{\Kpc}{~\mathrm{kpc}}
 
\newcommand{\Gyr}{~\mathrm{Gyr}}

\newcommand{\kmsec}{~\mathrm{km}~\mathrm{s}^{-1}}

\newcommand{\kmseckpc}{~\mathrm{km}~\mathrm{s}^{-1}~\mathrm{kpc}^{-1}}

\newcommand{\vc}{v_{\mathrm{c}}}

\newcommand{\de}{\mathrm{d}}

\newcommand{\Rg}{R_\mathrm{g}}

\newcommand{\RNum}[1]{\uppercase\expandafter{\romannumeral #1\relax}}

\newcommand{\degree}{^{\circ}}

\newcommand{\ldd}[2]{\de #1/\de #2}
\newcommand{\dd}[2]{\frac{\de #1}{\de #2}}
\newcommand{\pare}[1]{\left(#1\right)}
\newcommand{\paresq}[1]{\left[#1\right]}
\newcommand{\parec}[1]{\left\{#1\right\}}

\newcommand{\img}{\mathrm{i}}
\newcommand{\Rep}{\operatorname{Re}}
\newcommand{\Imp}{\operatorname{Im}}
\newcommand{\Eq}[1]{Eq.~(\ref{#1})}
\newcommand{\Eqs}[2]{Eqs.~(\ref{#1})-(\ref{#2})}

\newcommand{\Fig}[1]{Fig.~\ref{#1}}

\newcommand{\Phia}{\Phi_\mathrm{a}}

\newcommand{\Sec}[1]{Section~\ref{#1}}

\newcommand{\sgn}{\mathrm{sgn}}

\newcommand{\phib}{\phi_{\mathrm{b}}}
\newcommand{\phisp}{\phi_{\mathrm{sp}}}
\newcommand{\Js}{J_{\mathrm{s}}}
\newcommand{\Jf}{J_{\mathrm{f}}}

\newcommand{\ths}{\theta_{\mathrm{s}}}
\newcommand{\thf}{\theta_{\mathrm{f}}}
\newcommand{\oHp}{\overline{H'}}
\newcommand{\Vp}{V_\mathrm{p}}
\newcommand{\Vmax}{V_\mathrm{max}}
\newcommand{\Ep}{E_\mathrm{p}}

\newcommand{\Ej}{E_\mathrm{J}}
\newcommand{\alphab}{\alpha_\mathrm{b}}
\newcommand{\alphasp}{\alpha_\mathrm{sp}}
\newcommand{\Rsp}{R_\mathrm{sp}}
\newcommand{\calS}{{\cal S}_{jj'}}

\title[The dynamical signature of slow and fast bars]{Staying away from the bar: the local dynamical signature of slow and fast bars in the Milky Way} 
\author[G. Monari et al.]  {Giacomo~Monari$^{1}$\thanks{Email:~\texttt{giacomo.monari@astro.unistra.fr}},  Benoit~Famaey$^{1}$, Arnaud~Siebert$^{1}$, Aurore~Duchateau$^{1,2}$, \newauthor Thibault~Lorscheider$^{1,2}$ and Olivier~Bienaym\'e$^{1}$ \\ 
$^1$ Observatoire astronomique de Strasbourg, Universit\'e de Strasbourg, CNRS UMR 7550, 11 rue de
  l'Universit\'e, 67000 Strasbourg, France \\
$^2$ UFR de Math\'ematique et d'Informatique, Universit\'e de Strasbourg, 7 rue Ren\'e Descartes, 67084 Strasbourg, France
} \date{Released 2016}

\pagerange{\pageref{firstpage}--\pageref{lastpage}} \pubyear{2016}

\def\LaTeX{L\kern-.36em\raise.3ex\hbox{a}\kern-.15em
    T\kern-.1667em\lower.7ex\hbox{E}\kern-.125emX}

\begin{document}

\label{firstpage}

\maketitle

\begin{abstract}
  Both the three-dimensional density of red clump giants and the gas
  kinematics in the inner Galaxy indicate that the pattern speed of
  the Galactic bar could be much lower than previously
  estimated. Here, we show that such slow bar models are unable to
  reproduce the bimodality observed in local stellar velocity
  space. We do so by computing the response of stars in the Solar
  neighbourhood to the gravitational potential of slow and fast bars,
  in terms of their perturbed distribution function in action-angle
  space up to second order, as well as by identifying resonantly
  trapped orbits. We also check that the bimodality is unlikely to be
  produced through perturbations from spiral arms, and conclude that,
  contrary to gas kinematics, local stellar kinematics still favour a
  fast bar in the Milky Way, with a pattern speed of the order of
  almost twice (and no less than 1.8 times) the circular frequency at
  the Sun's position. This leaves open the question of the nature of
  the long flat extension of the bar in the Milky Way.
\end{abstract}


\section{Introduction}\label{sect:intro}

The Milky Way is a barred galaxy. This conclusion can be readily
established from the gas kinematics in the inner Galaxy
\citep[e.g.,][]{deVaucouleurs1964,Binney1991}, as well as from
near-infrared photometry \citep{Binney1997}. Nevertheless, and rather
surprisingly, the structural parameters of the Milky Way bar, in
particular its strength, orientation and pattern speed, are still very
poorly constrained.

Almost two decades ago, based on photometry and gas kinematics
arguments, a consensus emerged for a fast bar with corotation (CR)
around $\sim 3.5$~kpc \citep[e.g.,][]{Binney1997,Bissantz2003}, i.e. a
perturbation pattern speed $\Omb \approx 1.9 \, \Omega_0$ where
$\Omega_0$ is the local rotational frequency at the Sun's radius
$R_0$, and an angle between the bar major axis and the Galactic
centre-Sun direction of $\phib \sim 25\degree$. This pattern speed
would place the Sun just outside the outer Lindblad resonance (OLR) of
the bar, and the kinematical signature associated with this position
indeed appears to be present in the stellar phase-space distribution
in the Solar neighbourhood \citep[e.g.,][see also
  \Sec{sect:bim}]{Dehnen1999,Dehnen2000,Famaey2005,Minchev2007,Bovy2010,Quillen2011},
as well as possibly in large-scale stellar velocity fluctuations
\citep{Monari2014,Bovy2015}.

However, from the photometric point of view, the situation has
recently changed quite dramatically, since \cite{WeggGerhard2013} and
\cite{Wegg2015} measured the three--dimensional density of red clump
giants in the inner Galaxy by combining various recent photometric
surveys. They concluded that the Milky Way contains a central
box/peanut bulge \citep{Combes1990,Athanassoula2005} which is the
vertical extension of a longer, flatter bar, oriented at an angle of
$\phib \sim 27\degree$ from the Galactic centre-Sun direction, but
reaching out to a radius $\Rb \sim 5$~kpc. Since the bar cannot
physically extend beyond its corotation, this limits the pattern speed
of the bar. Simulated bars are usually rather shorter than their
corotation, and indeed by constructing dynamical models reproducing
this new bar density as well as the stellar kinematics from the BRAVA
survey, the pattern speed was estimated to be of the order of $\Omb
\approx \Omega_0$ \citep{Portail2015} placing the bar corotation very
near to the Sun. Two independent subsequent re-analyses of gas
kinematics in the inner Galaxy by \cite{Sormani2015} and \cite{Li2016}
then favored slightly higher pattern speeds, of the order of $\Omb
\approx 1.45 \, \Omega_0$ and $\Omb \approx 1.2 \, \Omega_0$
respectively, but both still much lower than the older estimate $\Omb
\approx 1.9 \, \Omega_0$.

Given this state of affairs, we now study in the present contribution
the effect of such low pattern speeds on stellar kinematics in the
Solar neighbourhood, which was previously considered a strong argument in favor
of a fast bar. In \Sec{sect:bim}, we briefly review the characteristic
observational signatures of non-axisymmetries in the Solar
neighbourhood, and more specifically the prominent Hercules moving
group in velocity space. We then review in \Sec{sec:analytical} the
expected theoretical form \citep[][hereafter M16]{Monari2016} of the
first order response of the phase-space distribution function (DF) in the
presence of non-axisymmetric potential perturbations. We then extend this
analysis up to second order, and also identify the location of resonantly
trapped orbits in local velocity space. We subsequently confront the
predictions to observations in the case of bars with low (\Sec{sect:low})
and high (\Sec{sect:high}) pattern speed. Conclusions are drawn in
\Sec{sect:concl}.

\section{The bimodal local velocity space}\label{sect:bim}

The study of the fine structure of stellar velocity space in the Solar
Neighbourhood dates back to more than a century and the work of
\cite{Proctor1869} and \cite{Kapteyn1905}, leading to the discovery of
the Hyades and Ursa Major moving groups. These are spatially unbound
groups of stars sharing similar velocities, but not the same age nor
chemical composition \citep[e.g.,][]{Famaey2005,Famaey2007,Famaey2008},
and hence most likely associated with perturbations by disc
non-axisymmetries. Another prominent structure in velocity space for
late-type stars was discovered by \cite{Eggen1958} and
\cite{Blaauw1970}: it is located further away from the centre of the
$UV$-space\footnote{Here we use the common notation of Galactic
  Astronomy in which, at the position of the Sun in the Galactic
  plane, $U$ is the stellar velocity towards the Galactic centre, and
  $V$ the velocity in the direction of Galactic rotation, both {\it
    with respect to the Sun}.} than the Hyades and Ursa Major groups,
with velocities similar to the $\xi$ Herculis star. This structure
creates a true secondary mode in local velocity space. This bimodal
velocity space is thus separated into a main (high-$V$) mode, and a
secondary (low-$V$) mode made up of this large moving group, referred
to as the Hercules stream, or Hercules moving group. Analysis of the
chemical abundances of this moving group revealed properties of a
mixed population of thin and thick disc stars, with a prevalence of
metal-rich thin disc stars, consistent with a dynamical perturbation
from a non-axisymmetry of the potential
\citep[e.g.,][Antoja~et~al.~2016 in
  preparation]{SoubiranGirard2005,Bensby2007,Ramya2016}. With a wavelet
analysis of the $UV$-plane based on Hipparcos astrometric data
combined with CORAVEL radial velocities for K and M giants
\citep{Famaey2005}, the structure has actually been identified by
\citet{Famaey2008} as a double-peak structure centred on $(U,V)
\simeq (-35,-51)\kmsec$ for the first `H1' peak, and $(U,V) \simeq
(-57,-51)\kmsec$ for the second `H2' peak.

The location of the Hercules moving group in velocity space has been
shown by \cite{Dehnen1999bar,Dehnen2000} and \cite{Muhlbauer2003} to
be a natural dynamical signature of the bar if the OLR radius is
located just inside the Solar position, since the orbits aligned with
the orientation of the bar at the Sun (the main mode) co-exist with
orbits anti-aligned with the bar \citep{Athanassoula1983}, which are
responsible for the Hercules stream, while unstable orbits have been
shown to be responsible for the observed gap between the main mode and
Hercules. This implies a pattern speed of the order of $\Omb \approx
1.9 \, \Omega_0$. With such a fast bar model, it was subsequently
shown that the Oort constants could be reproduced \citep{Minchev2007},
and that the expected azimuthal velocity location of Hercules as a
function of Galactocentric radius complied with data from the RAVE
survey \citep{Antoja2014}. The double-peak structure identified by
\cite{Famaey2008} within Hercules is however not reproduced by a
bar-only model, and has been suspected to be linked to another
perturbation, most probably spiral arms. Finally, another feature
which might presumably be associated with the bar is located at $(U,V)
\simeq (75,-55)\kmsec$ \citep{Dehnen1998}, denoted hereafter as the
`horn' of the velocity distribution.

Note that the precise location of Hercules in velocity space after
correction for the reflex Solar motion is slightly dependent on the
Solar motion itself. Considerable debate still exists regarding this
motion, especially in the $V$ direction. By extrapolating the
asymmetric drift relation to zero velocity dispersion,
\cite{DehnenLSR} estimated that the Sun moves in the direction of
Galactic rotation only slightly faster than the circular velocity,
namely $V_\odot = 5.25 \kmsec$, while they estimated $U_\odot =10
\kmsec$ by simply assuming no mean radial motion from the Local
Standard of rest itself. More recent discrepant values for the solar
motion include \cite{Schonrich2012} estimating $U_\odot =14 \kmsec$
and $V_\odot = 12 \kmsec$, as well as \cite{Bovy2015} estimating
$U_\odot =10 \kmsec$ and $V_\odot = 24 \kmsec$. Hereafter, we will
display the position of the two peaks (H1 and H2) of the Hercules
stream in peculiar velocity space $(u,v)$ for these different values
of the solar motion. The local circular speed $v_0$ and Sun's distance
from the centre of the Galaxy $R_0$ will be fixed such that $v_0 +
V_\odot = 30.24 \kmsec {\rm kpc}^{-1} \times R_0$ \citep{Reid2004}.
In the \cite{Schonrich2012} and \cite{Bovy2015} cases we use the
values for $v_0$ ($238$ and $218\kmsec$, respectively), and $R_0$
($8.27\Kpc$ and $8\Kpc$, respectively) proposed by the authors. In the
\cite{DehnenLSR} case we assume $R_0=8\Kpc$, which corresponds to
$v_0=236.67\kmsec$. Contrary to most previous studies, we will
concentrate on the form of the phase-space DF
expected in the Solar neighbourhood from perturbation theory, in the
spirit of M16, rather than on individual orbits.

\section{Perturbed distribution functions}\label{sec:analytical}
\subsection{General case: first order response}
Let $(R,\phi,z)$ be a cylindrical coordinate system with origin at the
centre of the Milky Way. The bulk of the mass of the Galaxy is
associated with an axisymmetric gravitational potential $\Phi_0(R,z)$,
and it is well-known that realistic galactic potentials are close to
integrable ones. In this case, the natural canonical coordinates for
dynamics are the action-angle variables \citep[see,
  e.g.,][]{BT2008,Fouvry2015}, $(\bJ,\bth)$, where $\bJ$ are integrals
of motion in $\Phi_0$. Thanks to the Jeans theorem, a stellar
population described by an axisymmetric phase-space distribution
function (DF) $f_0=f_0(\bJ)$ is in equilibrium.

Here, we follow the approach of M16 \citep[see
also][]{Kalnajs1971,CarlbergSellwood1985} in which a non-axisymmetric
perturbing potential $\Phi_1(R,\phi,z,t)$ is expanded in a Fourier
series of the angles $\bth$ as
\begin{equation}\label{eq:Phi1}
   \Phi_1(\bJ,\bth,t)= \Rep\parec{g(t)h(t)\sum_{\bn} c_{\bn}(\bJ) \eexp^{\img \bn \cdot \bth}},
\end{equation}
where $g(t)$ controls the growth of the perturbation with time, and
$h(t)$ is a periodic sinusoidal function, of frequency $\omp$, which
accounts for the perturbing potential rotating at a fixed pattern
speed. Typically, $\omp = -m \Omp$ where $m$ is the multiplicity (or
azimuthal wavenumber) of the perturbing potential and $\Omp$ its
pattern speed, and $h(t) = \exp(\img \omp t)$.

We assume that $g(t)$ is a well behaved function, that the
perturbation and its time derivatives were null far back in time
($g^{(k)}(-\infty)=0$), and that the perturbation has constant
amplitude at the present time ($g^{(0)}(t)=1$, and $g^{(k)}(t)=0$ for
$k=1,...,\infty$). Using the linearized Boltzmann equation, the linear
response to first order of the stellar equilibrium distribution $f_0$
to the perturbing potential $\Phi_1$ has been shown by M16 to be
$f=f_0+f_1$ with
\begin{equation}\label{eq:f1}
  f_1(\bJ,\bth,t)=\Rep\parec{\ddp{f_0}{\bJ}(\bJ)\cdot\sum_{\bn}\bn c_{\bn}(\bJ)
  \frac{h(t)\eexp^{\img\bn\cdot\bth}}{\bn\cdot\bom+\omp}}.
\end{equation}

The predictions of this perturbed DF have been directly compared to
test-particle simulations in M16, and showed remarkable agreement in
terms of the moments of the DF.  One might however wonder what is
gained by our analytical treatment of the problem compared to the
results of these test-particle simulations. The big difference is
that, contrary to such simulations, our analytical DFs will in the
future allow us to fit the data directly, with a few fitting
parameters in the perturbing potential as well as in the axisymmetric
DF, by performing a maximum-likelihood estimate of these parameters
based on actual kinematical data for a large set of individual stars.

The DFs computed from perturbation theory however diverge for resonant
orbits, for which $\bn\cdot\bom(\bJ)+\omp=0$. Resonances are
responsible for the `trapping' of orbits, which renders the former
linear treatment inappropriate near to resonances. We will identify
the resonantly trapped orbits in \Sec{sect:trapping}, in order to
display their location in local velocity space. But, first, we will
expand our perturbative treatment to second order, to make sure that
we are not missing any peculiar behavior of the DF with our first
order treatment of the problem.

\subsection{General case: second order response}\label{sec:analyticalf2}
In the previous section we derived the linear response $f_1$ of the DF
to the perturbing potential $\Phi_1$. `Linear' means that if
$|\Phi_1/\Phi_0|\sim \varepsilon \ll 1$, then $f_1 \in
\mathcal{O}(\varepsilon)$, and we neglect all higher order terms.  We
can expand the DF to the second order as
$f=f_0+f_1+f_2+\mathcal{O}(\varepsilon^3)$, where $f_2 \in
\mathcal{O}(\varepsilon^2)$. Plugging this expression in the
collisionless Boltzmann equation, and grouping together the
$\mathcal{O}(\varepsilon)$, $\mathcal{O}(\varepsilon^2)$, and
$\mathcal{O}(\varepsilon^3)$ terms, we obtain the linear equation for
$f_1$ that was used to obtain \Eq{eq:f1}, i.e.,
\begin{equation}
  \dd{f_1}{t}+[f_0,\Phi_1]=0,
\end{equation}
and the second order equation
\begin{equation}
  \dd{f_2}{t}+[f_1,\Phi_1]=0.
\end{equation}
Therefore,
\begin{equation}
  f_2(\bJ,\bth,t)=-\int_{-\infty}^t \de t~[f_1,\Phi_1].
\end{equation}
We can rewrite $f_2$ as $f_2=\tilde{f_2}-\hat{f_2}$, where
\begin{equation}\label{eq:f2t}
  \tilde{f_2}(\bJ,\bth,t)\equiv \int_{-\infty}^t \de t~\ddp{f_1}{\bJ}\cdot\ddp{\Phi_1}{\bth}, 
\end{equation}
and
\begin{equation}\label{eq:f2h}
  \hat{f_2}(\bJ,\bth,t)\equiv \int_{-\infty}^t \de t~\ddp{\Phi_1}{\bJ}\cdot\ddp{f_1}{\bth}.
\end{equation}

\subsection{Rotating bar case}
\subsubsection{First order}
As \cite{Weinberg1994} and \cite{Dehnen2000}, we assume that the
perturbing potential due to the Galactic bar behaves, outside from the
bar itself, as a quadrupole. Furthermore, we are only interested here
in the response inside the $z=0$ Galactic plane, i.e. we write for
the bar potential
\begin{equation}\label{eq:Fmode}
  \Phi_1(R,\phi,t)=\Rep\parec{\Phia(R)\eexp^{\img m(\phi - \phib -\Omb t)}},
\end{equation}
where $m=2$, $\Omb$ is the pattern speed of the perturber in the bar case,
$\phib$ is the angle between the Sun and the long axis of the bar, and
\begin{equation}
  \Phia(R)=-\alphab\frac{v_0^2}{3}\pare{\frac{R_0}{\Rb}}^3
  \times \left\{
  \begin{array}{l l}
    (R/\Rb)^{-3}  & \quad \text{} R \geq \Rb,\\
    2 -(R/\Rb)^{3} & \quad \text{} R < \Rb,
  \end{array} \right.
\end{equation}
where $(R,\phi)$ are the Galactocentric radius and azimuth, $\Rb$ is
the length of the bar, and $\alphab$ represents the maximum ratio
between the bar and axisymmetric background radial forces at the Sun's
position $R=R_0$ \citep[see also][]{Monari2015,Monari2016b}. 

Using \Eq{eq:f1}, and making use of the epicyclic approximation, M16
derived the form of $f_1$ for any rotating Fourier mode of the kind
\Eq{eq:Fmode}, for stellar populations close to the Galactic plane and
on low eccentricity orbits. For stars orbiting on the Galactic plane
(i.e. with $z=0$ and $v_z=0$), $f_1$ reads
\begin{equation}\label{eq:DF}
  f_1 = \Rep\Bigg\{\sum_{j=-1}^{1} c_{j m}
  {\rm F}_{j m}\eexp^{\img \paresq{j\theta_R+m\pare{\theta_\phi-\phib-\Omb t}}}\Bigg\},
\end{equation}
with
\begin{align}\label{eq:cR}
  c_{j m}(J_R,J_\phi)&\equiv \Bigg[\delta_{j 0}+\delta_{|j|1}\frac{m}{2}\sgn(j)
    \gamma e\Bigg]\Phia(\Rg,0) \nonumber \\
  &\quad -\delta_{|j|1}\frac{\Rg}{2}e\ddp{\Phia}{R}(\Rg,0),
\end{align}
and
\begin{equation}\label{eq:F}
  {\rm F}_{j m}(J_R,J_\phi)\equiv \frac{j\dddp{f_0}{J_R}+
    m\dddp{f_0}{J_\phi}}{j\kappa+m\pare{\omega_\phi-\Omb}}.
\end{equation}
Hereabove, we denote the circular and epicyclic frequencies with the
usual notation $\Omega$ and $\kappa$, which are evaluated at the
guiding radius $\Rg(J_\phi)$, defined as the radius where
$\Rg^2\Omega(\Rg)=J_\phi$. We also define
$\gamma\equiv2\Omega/\kappa$, the eccentricity
$e(J_R,J_\phi)\equiv\sqrt{2J_R/(\kappa\Rg^2)}$, and the azimuthal
frequency
$\omega_\phi(J_R,J_\phi)\equiv\Omega+(\ldd{\kappa}{J_\phi})J_R$.

Within the epicyclic approximation, we can relate the actions
$(J_R,J_\phi, \theta_R,\theta_\phi)$ to the usual cylindrical
phase--space coordinates $(R,\phi,v_R,v_\phi)$ through
\begin{equation}
\begin{aligned}
  J_R & =\frac{v_R^2}{2\kappa}+\frac{\kappa(R-\Rg)}{2},\quad & J_\phi&=R v_\phi, \\
  \theta_R &= \tan^{-1}\pare{-\frac{v_R}{\kappa (R-\Rg)}},\quad &
  \theta_\phi &= \phi+\Delta\phi,
\end{aligned}
\end{equation}
where 
\begin{equation}
  \Delta \phi\equiv -\frac{\gamma}{\Rg}\sqrt{\frac{2 J_R}{\kappa}}\sin\theta_R-
                    \frac{J_R}{2}\dd{\ln \kappa}{J_\phi}\sin(2\theta_R).
\end{equation}
Using the above relations, the phase-space DF
$f=f_0+f_1$ can be expressed as a function of the usual phase-space
coordinates, i.e. $f=f(R,\phi,v_R,v_\phi)$.

\subsubsection{Second order}\label{sect:sec_ord_bar}
We now compute the second order response $f_2$ from $f_1$ and $\Phi_1$
in the case case of a quadrupole potential like the one of
\Eq{eq:Fmode}. Once the derivatives in \Eqs{eq:f2t}{eq:f2h} are
calculated, one needs to solve simple integrals of sinusoidal
functions, in the same fashion as in the $f_1$ case. Then, after
defining the two-dimensional vectors $\bn \equiv (j,m)$, and $\bN
\equiv (j',m)$, \Eqs{eq:f2t}{eq:f2h} become,
\begin{equation}\label{eq:f2tbar}
  \tilde{f}_2(\bJ,\bth)=-\frac{1}{2}\sum_{j,j'=-1}^1
  \ddp{}{\bJ}\pare{c_{jm}F_{jm}}\cdot\bN c_{j'm}\Imp\parec{\calS^{-}},
\end{equation}
\begin{equation}\label{eq:f2hbar}
  \hat{f}_2(\bJ,\bth)=-\frac{1}{2}\sum_{j,j'=-1}^1 \bn
  c_{jm}F_{jm}\cdot\ddp{c_{j'm}}{\bJ}\Imp\parec{\calS^{+}},
\end{equation}
where
\begin{align}
\begin{split}
  \calS^{\pm}(\bJ,\bth)\equiv-\img &\Big[ 
  \frac{\eexp^{\img\paresq{\pare{\bn+\bN}\cdot\bth-2m\phib-2m\Omb
          t}}} {\pare{\bn+\bN}\cdot\bom-2m\Omb} \\
    &\pm\pare{1-\delta_{j,j'}}\frac{\eexp^{\img\pare{j-j'}\theta_R}}{\pare{j-j'}\kappa} \Big].
\end{split}
\end{align}

With these expressions, we can now calculate the response to the bar
up to second-order $f=f_0+f_1+f_2$, which we will plot in velocity
space at a given point in the Galactic plane.

Note that higher-order terms with $m>2$ would in principle also appear
in the expansion of the bar potential of \Eq{eq:Fmode} for more
complex bar shapes than a pure quadrupole. These additional terms
would have their own associated perturbed DF and are not considered
here. These DFs would in principle be of second or higher order
compared to the quadrupole amplitude. As we will show in the next
section, the second order effects computed hereabove for the
quadrupole case are subdominant indeed compared to the first order
response, and this should thus be the case for these additional terms
in the perturbing potential too.

\subsubsection{Resonantly trapped orbits}\label{sect:trapping}
It can immediately be seen from \Eq{eq:F} that the linear response
$f_1$ to the rotating perturbation $\Phi_1$ diverges for resonant
orbits whose frequencies $\kappa$ and $\omega_\phi$ are such that
\begin{equation}
  l\kappa+m(\omega_\phi-\Omb)=0,
\label{eq:resonance}
\end{equation}
with $l=0$ (corotation resonance), or $l=\pm 1$ (Lindblad
resonances). These are the first-order resonances.

Insight on the dynamics of a star near a resonance can be obtained
using a canonical transformation of coordinates defined by the type-2
generating function \citep[e.g.,][]{Weinberg1994}
\begin{equation}\label{eq:S}
  S=\paresq{l\theta_R+m\pare{\theta_\phi-\Omb t}}\Js+\theta_R\Jf.
\end{equation}
The new angles and actions $(\thf,\ths,\Jf,\Js)$ are related to the
old ones by
\begin{equation}\label{eq:can}
\begin{aligned}
  \ths & =l\theta_R+m\pare{\theta_\phi-\Omb t},\quad & J_\phi & =m \Js, \\
  \thf &= \theta_R, \quad & J_R &= l\Js+\Jf.
\end{aligned}
\end{equation}
In these new canonical coordinates, the motion is described by the Hamiltonian
\citep[e.g.][Appendix D.4.6]{BT2008}
\begin{equation}
  H'(\thf,\ths,\Jf,\Js)=H(\theta_R,\theta_\phi,J_R,J_\phi,t)+\ddp{S}{t},
\end{equation}
Since $\lddp{S}{t}=-m\Omb\Js$, and rewriting $H$ as a function of the
new coordinates $(\thf,\ths,\Jf,\Js)$, $H'$ reads
\begin{equation}
  H'=H_0+\Rep\parec{\sum_{j=-1}^1c_{jm}\eexp^{\img\paresq{(j-l)\thf+\ths}}}-m\Omb\Js,
\end{equation}
where $H_0=H_0(J_R,J_\phi)$ is the Hamiltonian of the unperturbed
axisymmetric system and the $c_{jm}(J_R,J_\phi)$ coefficients are the
Fourier coefficients from \Eq{eq:cR}. In this case, $J_R$ and $J_\phi$
have to be understood as functions of $(\Jf,\Js)$, as given by the
canonical transformation in \Eq{eq:can}\footnote{Notice that for any
  given orbit $H'$ takes a value that we can call $\Ej$, which is an
  integral of motion of the perturbed system. $\Ej$ is the energy in
  the frame of reference rotating with the perturbation, and is
  usually known as the `Jacobi integral'.}.

The angle $\ths$ is usually called `slow angle' because near a
resonance, for an orbit in the axisymmetric background potential, it
evolves very slowly by definition of the resonance in
\Eq{eq:resonance}, while $\thf$ is called the `fast angle'. Since
$\thf$ evolves much faster than $\ths$, we can average $H'$ along
$\thf$ \citep[the averaging principle,
  e.g.,][]{Arnold,Weinberg1994,BT2008}, to obtain
\begin{equation}
  \oHp=H_0(\Jf,\Js)-m\Omb\Js+\Rep\parec{c_{lm}(\Jf,\Js)\eexp^{\img \ths}}.
\end{equation}
Since $\dot{\Jf}=-\lddp{\oHp}{\thf}=0$, $\Jf$ is an integral of
motion, and the motion can be described only in the $(\ths,\Js)$ plane. We
can further write the Hamilton equations as,
\begin{equation}
\begin{aligned}
  \dot{\ths}&=\ddp{\oHp}{\Js}=\Oms+\Rep\parec{\ddp{c_{lm}}{\Js}\eexp^{\img \ths}}, \\
  \dot{\Js} &=-\ddp{\oHp}{\ths}=-\Rep\parec{\img c_{lm}\eexp^{\img \ths}},
\end{aligned}
\end{equation}
where $\Oms$ is the angular frequency associated to $\ths$ in the
unperturbed axisymmetric Hamiltonian in the rotating frame $H_0 -
\Omega_b J_\phi$. Dropping all the terms which are
$\mathcal{O}(\varepsilon^2)$, we get:
\begin{equation}
  \ddot{\ths} \approx \Rep\parec{\img\pare{\Oms\ddp{c_{lm}}{\Js}
      -\ddp{\Oms}{\Js}c_{lm}}\eexp^{\img\ths}}.
\label{eq:pend}
\end{equation}
Note that if we Taylor expand the function
$\Oms(\lddp{c_{lm}}{\Js})-(\lddp{\Oms}{\Js})c_{lm}$ about $\Js$ and
again drop all the terms that are $\mathcal{O}(\varepsilon^2)$,
\Eq{eq:pend} becomes a one-dimensional pendulum equation which can be
rewritten as $\ddot{\ths}=-\ldd{\Vp(\ths)}{\ths}$, where
\begin{equation}
  \Vp(\ths)=\Rep\parec{\pare{\ddp{\Oms}{\Js}c_{lm}-\Oms\ddp{c_{lm}}{\Js}}\eexp^{\img\ths}}
\end{equation}
is the pendulum potential and 
\begin{equation}
\Ep=\dot{\ths}^2/2+\Vp(\ths) 
\label{Ep}
\end{equation}
its energy. The maximum of the potential $\Vmax$ is 
\begin{equation}
  \Vmax=\left|\ddp{\Oms}{\Js}c_{lm}-\Oms\ddp{c_{lm}}{\Js}\right|.
\label{Vmax}
\end{equation}
The angle $\ths$ describes the precession angle of the orbit with
respect to the closed resonant orbit in the frame of reference
rotating with the perturbation, while $\thf$ is the motion of the star
along its orbit itself \citep{Weinberg1994,BT2008}. For $\Ep < \Vmax$
the angle $\ths$ librates back and forward between two values, around
the closed orbit, while for $\Ep>\Vmax$, $\ths$ circulates. Orbits
that have $\Ep<\Vmax$ are the orbits `trapped at the resonance', while
the circulating orbits with $\Ep\gg\Vmax$ can be fully described by
the perturbative treatment explained in the previous sections.

In the following we will display the zone of local velocity space
corresponding to orbits trapped by the first-order resonances. This
trapping will affect the actual density of stars in velocity space in
the trapped zone compared to our models, but will not strongly affect
the general shape of velocity space itself.

\section{Slow bar models}\label{sect:low}
We will now explore the actual response of stars
in the solar neighbourhood to a bar perturbation.

\begin{figure*}
  \centering 
  \includegraphics[width=0.75\textwidth]{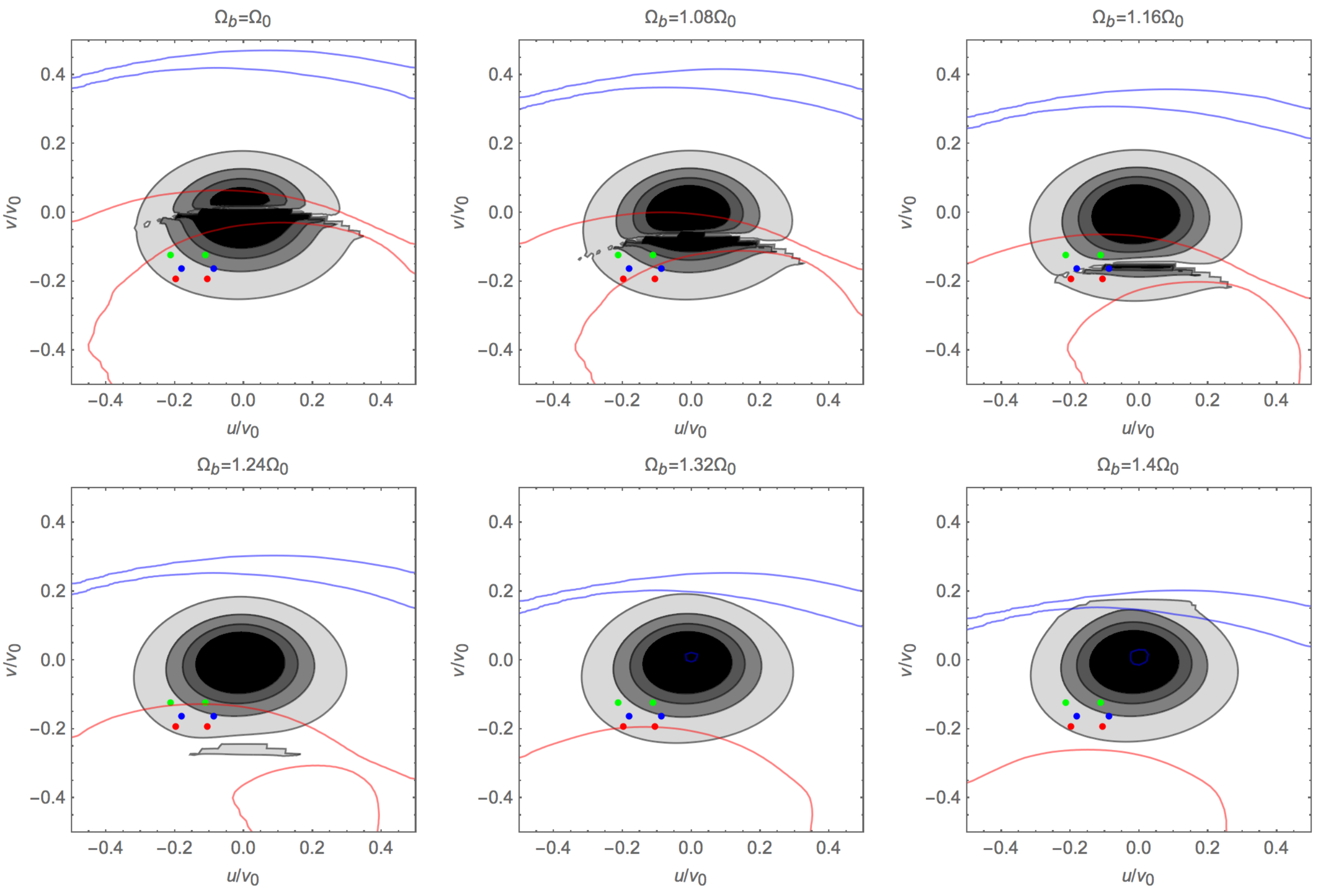}
  \caption{The local stellar velocity distribution perturbed to the
    linear order (i.e. the phase-space DF $f=f_0+f_1$, with $f_0$ the
    axisymmetric DF of M16 -- namely the quasi-isothermal DF first introduced
by \citet{Binney2011} as a modification of the DF defined by \citet{Binney2010}-- and $f_1$ given by \Eq{eq:DF} in the
    $(u,v)$ plane at $(R,\phi,z)=(R_0,0,0)$ for different bar models
    with low pattern speed ($\Omega_0 \leq \Omb \leq 1.4
    \Omega_0$). In this and all the figures of this work, the contours
    include 34, 50, 68, and 90 per cent of the stars
    respectively. Here the local maximum ratio at $R_0$ between the
    bar and axisymmetric background radial forces is
    $\alphab=0.01$. The colored points represent the {\it observed}
    Hercules moving group peaks H1 and H2 as estimated by
    \protect\cite{Famaey2008}, corrected for the Sun's motion in the
    estimates of \protect\cite{DehnenLSR} (red points),
    \protect\cite{Schonrich2012} (blue points), and
    \protect\cite{Bovy2015} (green points). The red (blue) contours
    delimit the region of resonant trapping by the CR (OLR),
    i.e. orbits with $\Ep<\Vmax$ as defined in \Eq{Ep} and
    \Eq{Vmax}. At $\Omb \simeq 1.45 \Omega_0$ \citep{Sormani2015}, the
    zone of influence of the CR (red contours) is clearly moving away
    from the bulk of stars in velocity space, whilst the OLR (blue
    contours) only has an influence at high $v$, far from the actual
    location of the Hercules moving group in local velocity space
    (coloured points). For $\Omb \simeq 1.2\Omega_0$ \citep{Li2016},
    the CR of the bar does create a bimodality at the right location
    in $v$, but not in $u$.}
  \label{fig:vRvphi_Wegg}
\end{figure*}
\begin{figure*}
  \centering 
  \includegraphics[width=0.75\textwidth]{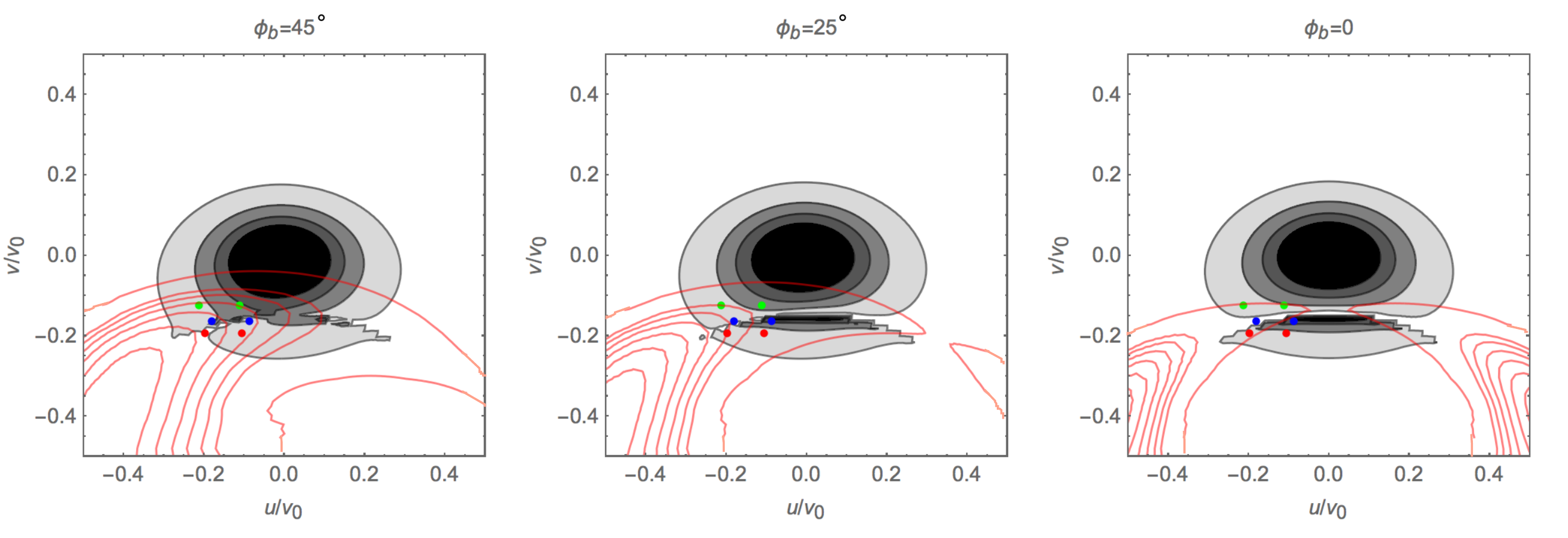}
  \caption{As in \Fig{fig:vRvphi_Wegg}, but this time fixing the
    pattern speed to $\Omb=1.16\Omega_0$, and varying the angle
    $\phib$. This does not help, as to get the bimodality at the right
    location to reproduce Hercules (colored points), one would need a
    negative $\phib$ clearly excluded from the photometric structure
    of the Milky Way bar. The inner red contours represent different
    values of $\eta=\Ep/\Vmax$, for orbits trapped at the bar's CR. In
    particular, they correspond to $\eta_i=\etamin+i\Delta\eta$, for
    $i=1,...,6$, $\etamin$ the minimum of $\eta$, and
    $\Delta\eta=(1-\eta)/10$. The trapping onto closed orbits
    (smallest $\eta$) is far from the bulk of stars in velocity
    space.}
  \label{fig:vRvphi_Wegg_ang}
\end{figure*}
Let $R_0$ be the Sun's Galactocentric radius and
$\Omega_0=\Omega(R_0)$ the local rotation frequency. As outlined in
\Sec{sect:intro}, recent models of the distribution and kinematics of
stars the inner part of the Milky Way
\citep{Long2013,Portail2015,Wegg2015} as well as models of the gas
kinematics \citep{Sormani2015,Li2016} favour pattern speeds estimates
of the bar much lower ($\Omega_0 \lesssim \Omb \lesssim 1.45
\Omega_0$) than the ones that for more than a decade were considered
as settled ($1.8 \Omega_0 \lesssim \Omb \lesssim 2 \Omega_0$).  While
the old estimates predict the OLR to be in proximity of the Solar
neighbourhood, the new estimates rather predict that stars in the
Sun's vicinity would be influenced by the bar's corotation (CR), or at
least that the Sun is located well inside the OLR. In this Section we
analyze what this would mean in terms of the shape of the DF in the
Solar neighbourhood.

Using a common notation in Galactic Astronomy, for each point of the
Galactic plane $(R,\phi)$, we define the peculiar velocity
coordinates\footnote{Note that in general, when we consider volumes of
  finite non-zero size, $u\neq- v_R$ and $v \neq v_\phi-\vc(R)$. In
  fact, $(u,v)$ in Galactic Astronomy refer to the Cartesian
  velocities in the Local Standard of Rest \citep{BT2008}. However, in
  the case of velocity space at a given point in configuration space,
  like we consider in this work, $u=-v_R$ and $v=v_\phi-\vc(R)$ is a
  good approximation.} of stars, $u\equiv-v_R$ and $v\equiv
v_\phi-\vc(R)$, where the circular velocity $\vc(R)\equiv
R\Omega(R)$. As in M16, we choose a quasi-isothermal DF for $f_0(\bJ)$, first introduced
by \citet{Binney2011} as a modification of the DF defined by \citet{Binney2010}, and reducing
to a Schwarzschild DF in the epicyclic approximation (see M16 for the details of the DF parameters).

Hereafter, when our perturbed $f_1$ (and $f_2$) diverges due to our
treatment to only 1st order (or 2nd order), we simply choose to set
$f_1=\sgn(f_1)|f_0|$ ($f_2=\sgn(f_2)|f_0|$) whenever $f_0<|f_1|$
($f_0<|f_2|$) due to $|f_1|$ (or $|f_2|$) being larger than $f_0$ near
first-order or second-order resonances. A detailed treatment of the
behaviour of the perturbed DF in the trapped regions will be the topic
of further work, but we already clearly identify the regions affected
by first-order resonant trapping thanks to the formalism developed in
\Sec{sect:trapping}. The regions of trapping at the resonances, where
$\Ep < \Vmax$ (see \Sec{sect:trapping}), will be surrounded on the
figures by red contours in the CR case, and by blue contours in the
OLR case.

\subsection{First order response in bar-only models}

In \Fig{fig:vRvphi_Wegg} we plot the value of the distribution
function $f=f_0+f_1$ as a function of the velocities $(u,v)$ expressed
as a function of the local circular speed $v_0$, for stars at
$(R,\phi)=(R_0,0)$ and $v_z=0$, i.e. orbiting on the Galactic plane
and passing in the Solar neighbourhood at the present time. We
consider a range of pattern speeds coherent with the recent estimates
by \cite{Portail2015}, \cite{Sormani2015} and \cite{Li2016}, namely
$\Omb=[\Omega_0,1.45\Omega_0]$.

To obtain this figure, we used the parameters $\phib=25\degree$ (which
controls the angle between the bar's long axis and the line connecting
the Sun and the Galactic centre), and $\alphab=0.01$. We set $\Rb =
0.625 R_0$, but also tried a case where the bulk of the mass is in the
inner parts of the bar, $\Rb = 0.44 R_0$, and found qualitatively
exactly the same patterns in velocity space.

As can be seen from the trapped regions in \Fig{fig:vRvphi_Wegg}, for
$\Omb<1.27\Omega_0$, the DF is not influenced by the OLR. The CR has
instead an influence for $v<0$: the velocity distribution is clearly
split into two parts that we call the `low--$v$' and `high--$v$'
modes. Stars in the high--$v$ mode have $u\lesssim 0$ (i.e. they have
the slight tendency to move outwards in the Galaxy), while stars in
the low--$v$ mode have $u\gtrsim 0$. The observed H1 and H2 velocity
peaks of the Hercules moving group, corrected for the Sun's motion
according to different estimates (see \Sec{sect:bim}), are represented
by the red points for the estimate of \cite{DehnenLSR}, by the blue
points for the estimate of \cite{Schonrich2012}, and by the green
points for the \cite{Bovy2015} estimate. As is evident from this
figure, even if the low--$v$ mode is formed because of the CR, for
none of the Sun's motion estimates is its position compatible with the
actual position of Hercules. Actually, the stars of the Hercules
moving group have on average $u<0$, while the low-$v$ mode generated
by the CR has $u\gtrsim 0$. On the other hand, when the pattern speed
increases to $\Omb\geq 1.27\Omega_0$ the OLR has a slight influence
only to a few stars with very high-$v$, not at all in the relevant
region of velocity space for the range of pattern speed considered in
this section.

Observations clearly show that $\phib\geq 0$
\citep{Binney1997}. However, a definitive estimate of the bar's angle
is still missing. In \Fig{fig:vRvphi_Wegg_ang} we thus explore the
effect of varying the bar angle in the range
$\phib=[0,45\degree]$. While, because of the symmetry of the model,
for $\phib=0$ the DF is exactly symmetric with respect to $u=0$,
increasing $\phib$ moves the low--$v$ mode to larger $u$, thus
increasing its distance from the actual position of the Hercules
moving group, and making the situation worse.

An alternative possibility would nevertheless be that the region of
strongest trapping, by construction not well taken into account by our
linear model, would correspond to the location of Hercules. We thus
also plot on \Fig{fig:vRvphi_Wegg_ang} the isocontours of the energy
of the pendulum $\Ep$ (see \Sec{sect:trapping}): we can see that the
region of strongest trapping close to the resonant closed orbit is outside of the bulk of stars in
velocity space, and would correspond to stars orbiting in the plane
but with extremely high eccentricity.

No feature in our modelled local velocity space can thus account for
the Hercules stream to first order in bar-only models.

\subsection{Second order response in bar-only models}

\begin{figure}
  \centering 
  \includegraphics[width=0.25\textwidth]{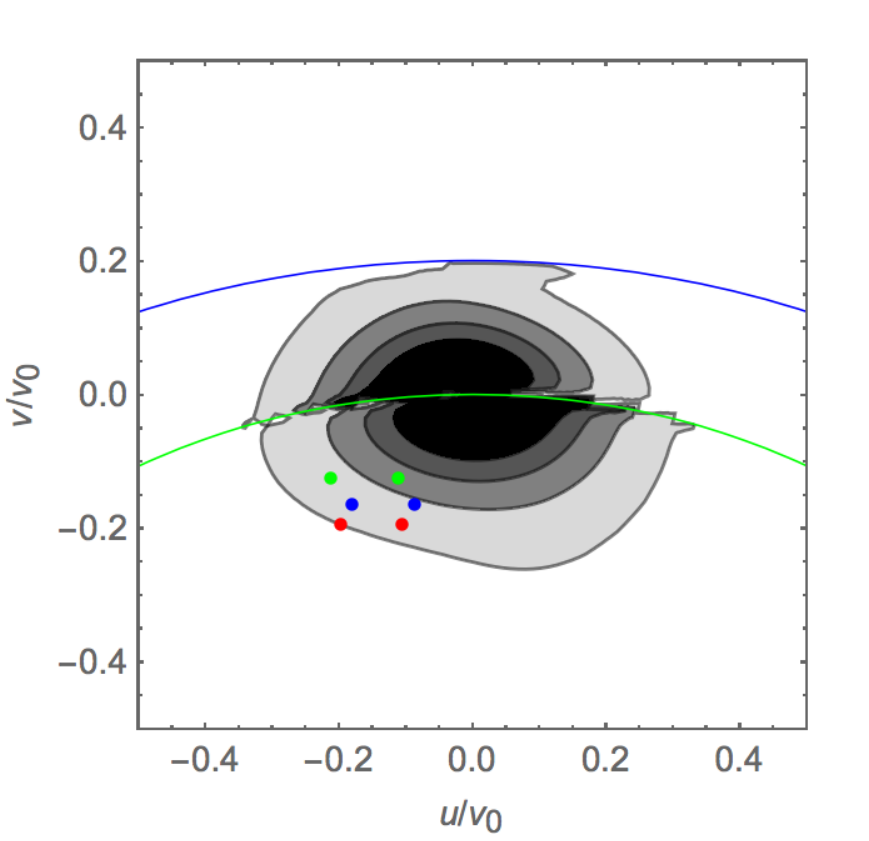}
  \caption{Illustration of the effect of the second-order response to
    the bar potential in an extreme unrealistic case with
    $\alphab=0.1$. The figure displays the velocity distribution in
    the $(u,v)$ plane at $(R,\phi,z)=(R_0,0,0)$ including only the
    second order DF response (i.e. $f=f_0+f_2$) for
    $\Omb=1.36\Omega_0$. The blue line displays the location of the
    OLR, and the green line the outer ultraharmonic resonance
    (OUHR). Here one can see that not only is the effect negligible
    for a realistic bar strength (i.e., in reality, one would have
    $\alphab \ll 1$), the location of the effect in velocity space
    also does not correspond to the Hercules moving group.}
  \label{fig:UH}
\end{figure}
\Fig{fig:UH} shows $f$ when $\Omb$ is chosen so that $R_0$ is at the
radius of the outer ultraharmonic resonance $\ROUH$, defined as the
radius $R$ where $\kappa(\ROUH)+4[\Omega(\ROUH)-\Omb]=0$. Near to
$\ROUH$ we have the strongest effect of the outer ultraharmonic
resonance,
\begin{equation}
  \kappa+4(\omega_\phi-\Omb)=0.
\end{equation}
The second order expansion of the collisionless Boltzmann equation
(\Sec{sect:sec_ord_bar}) takes into account the ultraharmonic
resonances, as appears clearly in the denominators of
\Eqs{eq:f2tbar}{eq:f2hbar}. For amplitudes of the bar potential
similar to those that we used in the previous Sections, the effects of
the second order resonance are small, and not enough to appreciate the
effects of the outer ultraharmonic resonance on the distribution
function $f$. Therefore, in \Fig{fig:UH} we plot an extreme case,
where the amplitude of the bar radial force is as large as 10 percent
of that of the axisymmetric background ($\alphab=0.1$), and we plot
only the effects of $f_2$. This extreme case is fully unrealistic for
the Milky Way, but clearly shows that the effect of the ultraharmonic
resonance of the bar cannot explain the formation of the Hercules
moving group: while $f$ is split in two parts in $v$, the low-$v$ mode
of the distribution has in average positive $u$, again contrary to the
observed behaviour of the Hercules moving group.

\subsection{Models with spiral arms}
\begin{figure*}
  \centering 
  \includegraphics[width=\textwidth]{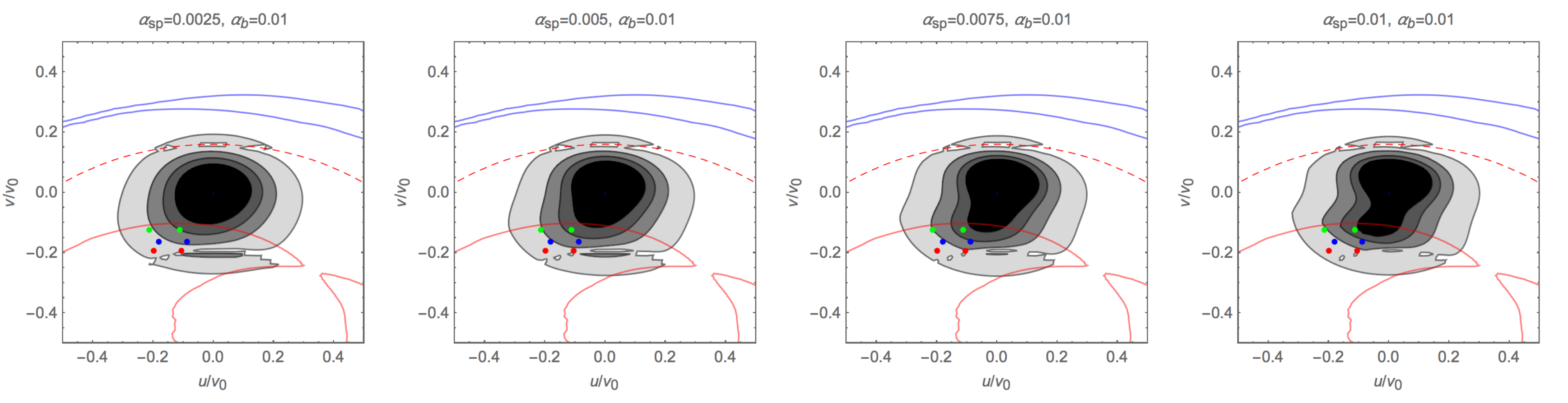}
  \caption{Linear response of the DF (i.e. $f=f_0+f_1$) in the $(u,v)$ plane
    at $(R,\phi,z)=(R_0,0,0)$, for a model with {\it both}
    a slow bar and a spiral pattern
   similar to the one of \protect\cite{Li2016}. It includes two $m=2$ spiral arms modes
    rotating with pattern speed $\Omsp=0.84\Omega_0$, and a bar with
    $\Omb=1.2\Omega_0$ and $\phib=27\degree$. Different panels display the response for
    different amplitudes of the spiral arms. The red (blue) solid
    contours delimit the regions of resonant trapping to the bar's
    CR (OLR). The red dashed line represents the CR of the spiral arms.}
  \label{fig:Li2sp}
\end{figure*}
We now analyze the combined effects of the bar potential and a spiral
pattern potential. Away from resonance overlap regions, the combined
linear response $f_1$ is then simply the sum of the response due to
the bar and that of the spiral arms (calculated in the same way as for
the bar, but for a different $\Phia$).

A question is whether spiral arms could be responsible for the
Hercules moving group. In their work, \cite{Li2016} propose a best-fit
model of the kinematics of the Milky Way, with a perturbing potential
composed by a slow bar ($\Omb=33\kmseckpc$), and two $m=2$ spiral arms
patterns, displaced by $\Delta\phisp=20.25\degree$ in azimuth,
rotating with an angular frequency $\Omsp=23\kmseckpc$.

We reproduce a spiral potential similar to that of \cite{Li2016} by
considering two Fourier modes with $m=2$, as in \Eq{eq:Fmode}. In this
case $\Omp=\Omsp=0.84\Omega_0$, and
\begin{equation}\label{eq:sp}
  \Phia(R)=\alphasp\Phi_0(R_0,0)\eexp^{-\img m \ln (R/\Rsp)/\tan p},
\end{equation}
where $\Rs=0.125R_0$. As a reference value we assume $\alphasp=0.005$
\citep{Siebert2012}. The locus of the arms and the pitch angle $p$ are
also like in \cite{Li2016}. The pattern speed of the bar is set at
$\Omb=1.2\Omega_0$, and $\Rb=0.625 R_0$. We show the results of this
model to the linear order in \Fig{fig:Li2sp}, for different values of
the spiral arm amplitude. It is clear from this figure that, up to the
linear order response, this model does not describe in a satisfying
way the kinematics of the Solar neighbourhood. 

\begin{figure}
  \centering 
  \includegraphics[width=\columnwidth]{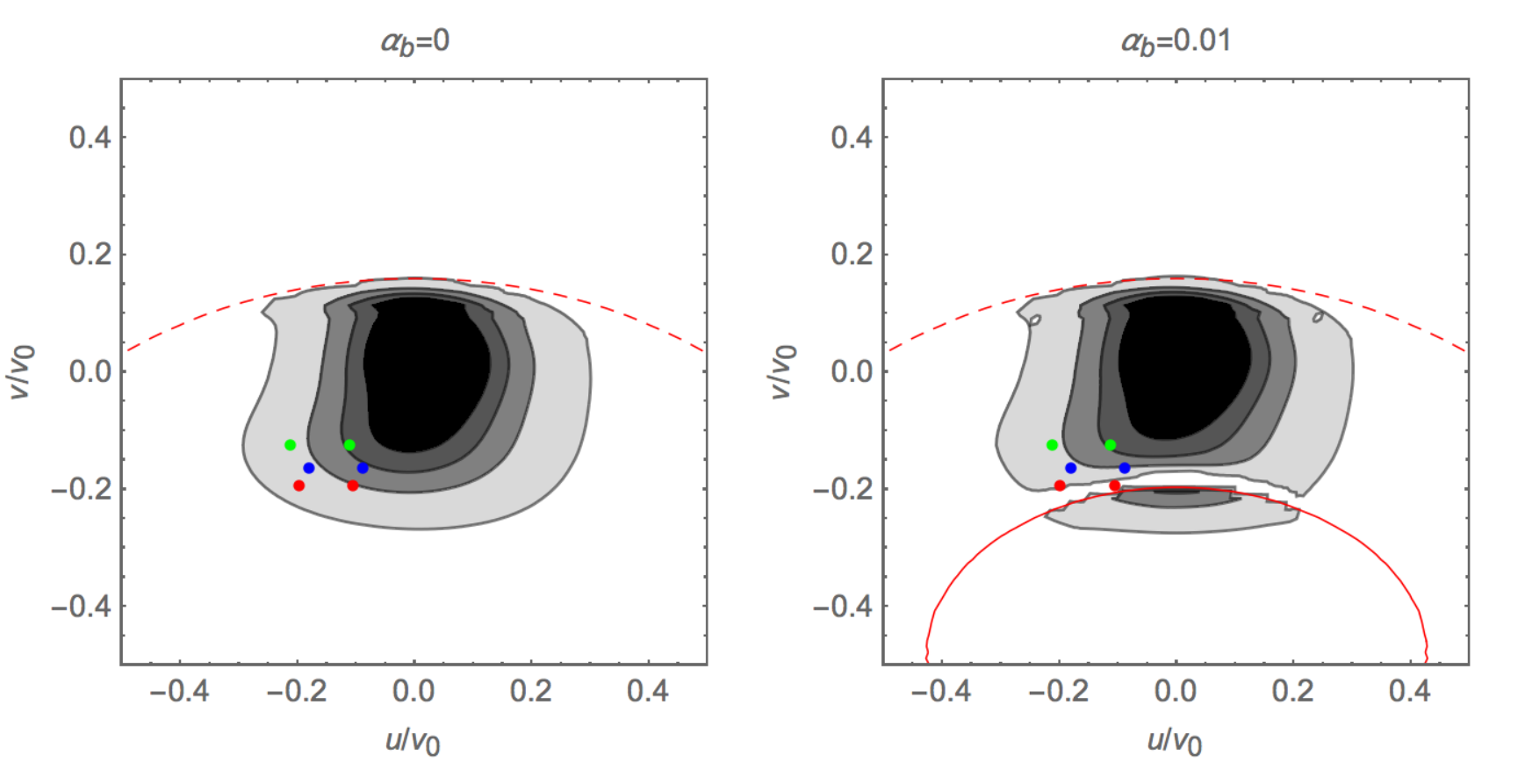}
  \caption{Effects of the second order response ($f=f_0+f_1+f_2$) on
    the model shown in \Fig{fig:Li2sp} for $\alphasp=0.005$, without
    (left panel) and with (right panel) the bar. The red solid
    (dashed) line represents the CR of the bar (spiral arms).}
  \label{fig:Gerhard2sp_2ndorder}
\end{figure}

In \Fig{fig:Gerhard2sp_2ndorder} we consider the second order
response, both to the spiral arms and bar perturbation.  The
calculation of $f_2$ in the case of the spiral arms is more cumbersome
than in the bar case, since the Fourier coefficients $c_{jm}$ are in
this case complex numbers, quadrupling the number of terms of
$f_2$. The $\tilde{f_2}$ and $\hat{f_2}$ components of
\Eqs{eq:f2t}{eq:f2h} become
\begin{align}\label{eq:f2tsp}
\begin{split}
  \tilde{f}_2(\bJ,\bth)=&-\sum_{j,j'=-1}^1\frac{\bN}{2}\cdot\Big[ \\
  &+\ddp{}{\bJ}\Rep\parec{c_{jm}F_{jm}}\Rep\parec{c_{j'm}}\Imp\parec{\calS^{-}} \\ 
  &+\ddp{}{\bJ}\Rep\parec{c_{jm}F_{jm}}\Imp\parec{c_{j'm}}\Rep\parec{\calS^{+}} \\
  &+\ddp{}{\bJ}\Imp\parec{c_{jm}F_{jm}}\Rep\parec{c_{j'm}}\Rep\parec{\calS^{-}} \\
  &-\ddp{}{\bJ}\Imp\parec{c_{jm}F_{jm}}\Imp\parec{c_{j'm}}\Imp\parec{\calS^{+}} \Big],
\end{split}
\end{align}
\begin{align}\label{eq:f2hsp}
\begin{split}
  \hat{f}_2(\bJ,\bth)=&-\sum_{j,j'=-1}^1\frac{\bn}{2}\cdot\Big[\\
  &+\Rep\parec{c_{jm}F_{jm}}\ddp{}{\bJ}\Rep\parec{c_{j'm}}\Imp\parec{\calS^{+}} \\
  &+\Imp\parec{c_{jm}F_{jm}}\ddp{}{\bJ}\Rep\parec{c_{j'm}}\Rep\parec{\calS^{-}} \\
  &+\Rep\parec{c_{jm}F_{jm}}\ddp{}{\bJ}\Imp\parec{c_{j'm}}\Rep\parec{\calS^{+}} \\
  &-\Imp\parec{c_{jm}F_{jm}}\ddp{}{\bJ}\Imp\parec{c_{j'm}}\Imp\parec{\calS^{-}} \Big].
\end{split}
\end{align}
The result is qualitatively fully unchanged, the second order response
of both the bar and spirals being very small in the region of
interest. One can see that the spiral indeed distorts the velocity
distribution somewhat in the right direction, but the effect is very
limited and does not at all render the velocity space bimodal as
observed. This slight distortion of velocity space is more likely
related to the observed Hyades and Ursa Major moving groups than to
Hercules.

\section{Fast bar models}\label{sect:high}
\begin{figure*}
  \centering 
  \includegraphics[width=\textwidth]{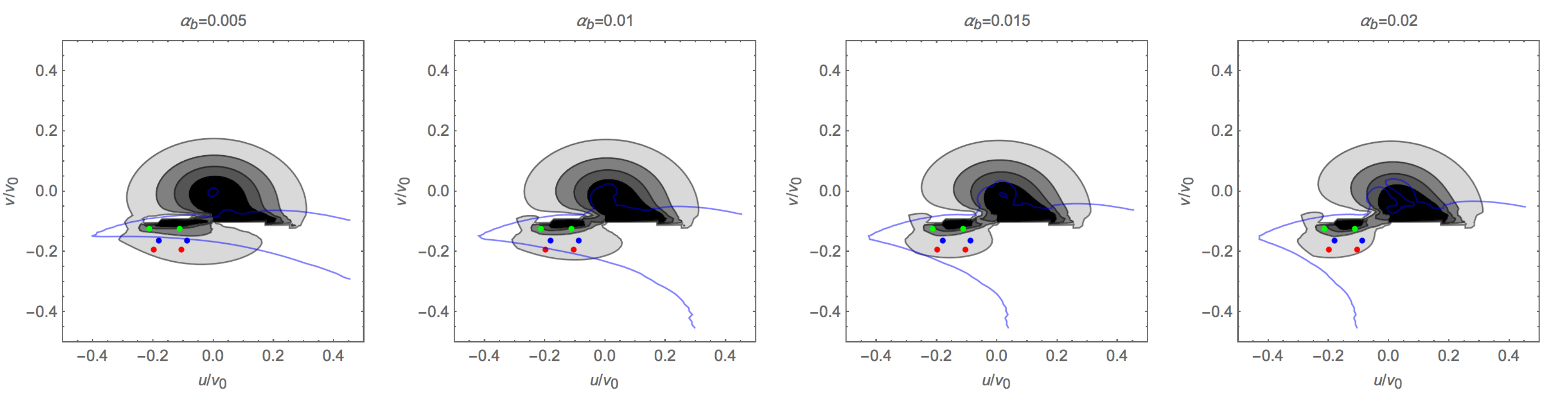}
  \caption{Velocity distribution to the second order
    (i.e. $f=f_0+f_1+f_2$) in the $(u,v)$ plane at $(R,\phi,z)=(R_0,0,0)$
    for a fast bar model $\Omb=1.89\Omega_0$ and $\phib=25\degree$. We
    consider different bar's amplitudes $\alphab$. The blue contours
    represent regions of trapping to the bar's OLR.}
  \label{fig:vRvphi_def}
\end{figure*}
In the models of this Section we show how our models reproduce the
traditional results expected in the Solar neighbourhood for a fast
bar. We use $\Omb=1.89\Omega_0$, following the estimates of
\cite{Antoja2014}. This pattern speed corresponds to $\ROLR=0.9R_0$,
where $\ROLR$ is defined as $R$ satisfying
\begin{equation}\label{eq:OLR}
  2\left[\Omega\pare{\ROLR}-\Omb \right ]+\kappa\pare{\ROLR}=0.
\end{equation}
In \Fig{fig:vRvphi_def} we show the results for $\phib=25\degree$ and
different values of the amplitude of the perturbation in the range
$\alpha=[0.005,0.02]$. The results keep in the account both the linear
and the quadratic response of the disc DF, $f_1$ and $f_2$
respectively. In this case, $f$ is split in two parts as observed in
\Sec{sect:bim}. Stars in the low-$v$ (high-$v$) mode have
$\omega_\phi<\Omb-\kappa/2$ ($\omega_\phi>\Omb-\kappa/2$), and tend to
move outwards (inwards) in the Galaxy, i.e. they have $u<0$
($u>0$). This is a direct consequence of the linear effects of the bar
on the stars' orbits. In particular, the orbits of stars with guiding
centre inside (outside) the OLR become elongated perpendicularly
(along) the bar \citep{BT2008}. \Fig{fig:vRvphi_def} in agreement with
the results of several of the numerical simulations of the effects of
the bar in the Solar neighbourhood, starting from the pioneering work
of \cite{Dehnen2000}. The stars forming the low-$v$ velocity mode are
usually associated with the Hercules moving group, and the gap between
this moving group and the main velocity mode in the Solar
neighbourhood were used by \cite{Antoja2014} to estimate the pattern
speed of the bar, assuming that the gap is due to the bar's OLR.

Increasing the bar's strength moves the stars closer to the resonance
curve. Moreover, stars in the low-$v$ (high-$v$) mode travel faster
outwards (inwards) in the Galaxy.  Interestingly, we see that while
all the estimates are compatible with the identification of the low
$v$ mode with the Hercules moving group, the estimates with higher
Sun's tangential velocity $V_\odot$ \citep{Bovy2015} is globally
favoured by the fast bar models considered here, and also favour
models with a stronger bar.

\begin{figure*}
  \centering 
  \includegraphics[width=0.75\textwidth]{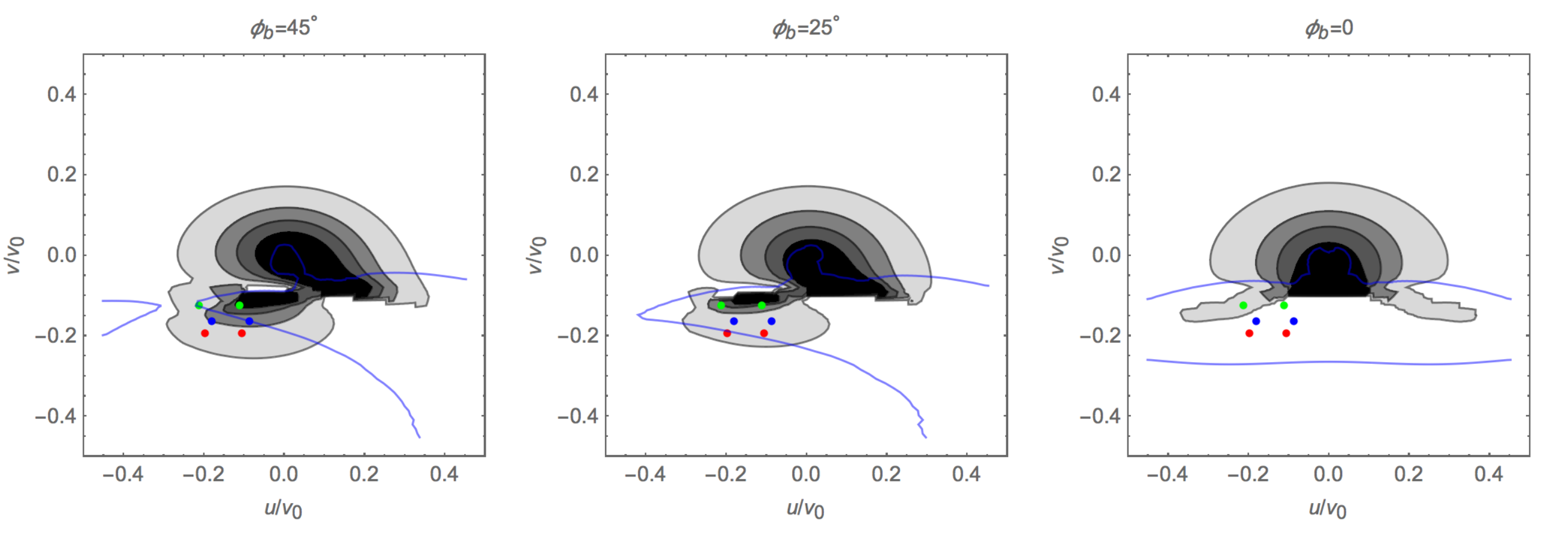}
  \caption{As in \Fig{fig:vRvphi_def}, but keeping fixed the bar's
    amplitude to $\alphab=0.01$, and varying the angle $\phib$.}
  \label{fig:vRvphi_ang}
\end{figure*}
In \Fig{fig:vRvphi_ang} we show the variation of the DF as a function
of $\phib$, keeping constant $R$. Also in this case the second order
effects are taken into account. We present only positive values of
$\phib$ as observations show that this is the case in the Milky Way
\citep{Binney1997}. Moreover, because of the symmetry of the DF in
\Eq{eq:DF}, the case of negative angles is readily obtained simply
flipping the signs of the $u$ velocities in \Fig{fig:vRvphi_ang}. At
$\phi=0$ the DF is completely symmetric with the respect of $u=0$,
while increasing the angle $\phib$ increases the number of stars with
negative (positive) $u$ for the low (high) $v$ velocity mode.

\begin{figure}
  \centering 
  \includegraphics[width=\columnwidth]{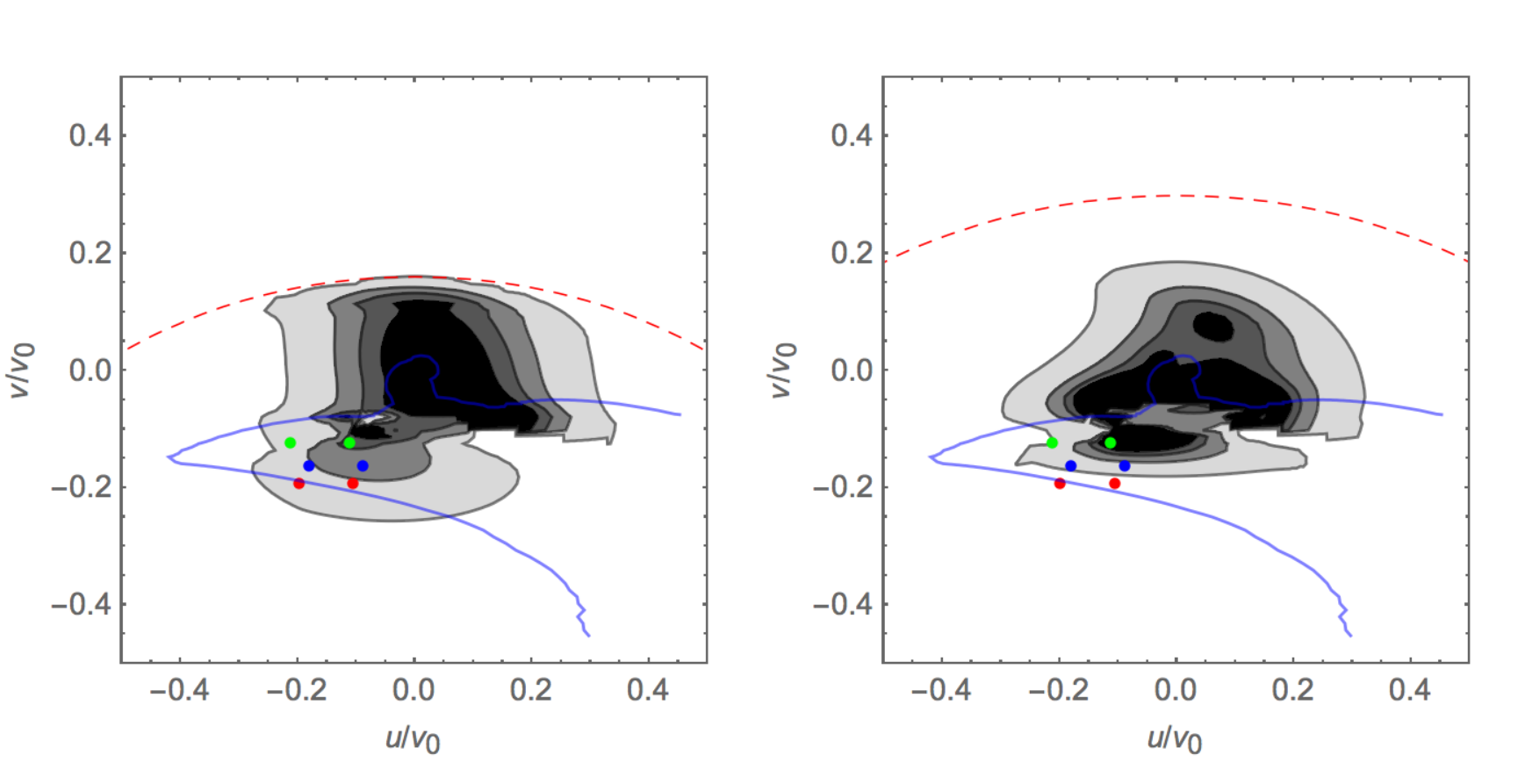}
  \caption{Second order response of the DF (i.e. $f=f_0+f_1+f_2$) for
    the spiral arms models of \protect\cite{Li2016} (left panel) and
    \protect\cite{Siebert2012} (right panel), and a fast bar with
    $\Omb=1.89\Omega_0$ and $\phib=25\degree$ in the $(u,v)$ plane at
    $(R,\phi,z)=(R_0,0,0)$.}
  \label{fig:vRvphi_sp_Dehn}
\end{figure}
Then, using the form of the spiral arms potential in
\Eq{eq:sp}, we study the combined effects of the fast bar and spiral
arms up to the second order. 

In \Fig{fig:vRvphi_sp_Dehn} we show $f$ at $(R,\phi)=(R_0,0)$ in the
plane for the combination of the fast bar and two spiral arms
models. The figures keep in account the first and second order
effects, both for the bar and the spiral arms. As bar parameters we
use $\Omb=1.89\Omega_0$, $\phib=25\degree$, and $\alphab=0.01$. The
first spiral arm model is formed by the two $m=2$ modes with the same
locus of the spiral arms as \cite{Li2016}. The second model is only
one $m=2$ mode, representing the old stellar arms, with the parameters
taken from \cite{Siebert2012}. The pattern speed is in the former case
$\Omsp=0.84\Omega_0$ \citep{Li2016}, and in the latter
$\Omsp=0.69\Omega_0$ \citep{Siebert2012}.

The comparison with \Fig{fig:vRvphi_sp_Dehn} shows that the
\cite{Siebert2012} spiral arms increase the probability to find stars
in the low-$v$ mode, and deforms the shape of both the low and
high-$v$ mode, and in particular the latter. The deformation in the
high-$v$ mode slightly resembles the Hyades moving group overdensity
in the Solar neighbourhood. In the same region of the velocity space
we can notice the effect of the inner ultraharmonic resonance, that
creates a small gap in the high-$v$ mode. Several authors suggested
\citep[e.g.,][]{Quillen2005,Pompeia2011} that the inner ultraharmonic
resonance of the spiral arms could be the cause of the Hyades moving
group. Our treatment seems to suggest that this resonance would be too
weak to influence alone the velocity distribution in such a
significant way. However, given its vicinity with the bar OLR, it
could be that the resonance overlap would render our treatment not
appropriate, and the coupling effects important
\citep[e.g.,][]{Monari2016b}. A fully proper treatment near
resonances, including resonance overlaps, will be the topic of further
work. We however note that the simulations of \citet{Monari2016b} did
not show significant differences for the in-plane motions between the
coupled simulation and the linear combination of simulations with a
single perturber. The difference was much more pronounced in terms of
vertical motions, which we do not consider here.

\begin{figure}
  \centering 
  \includegraphics[width=0.25\textwidth]{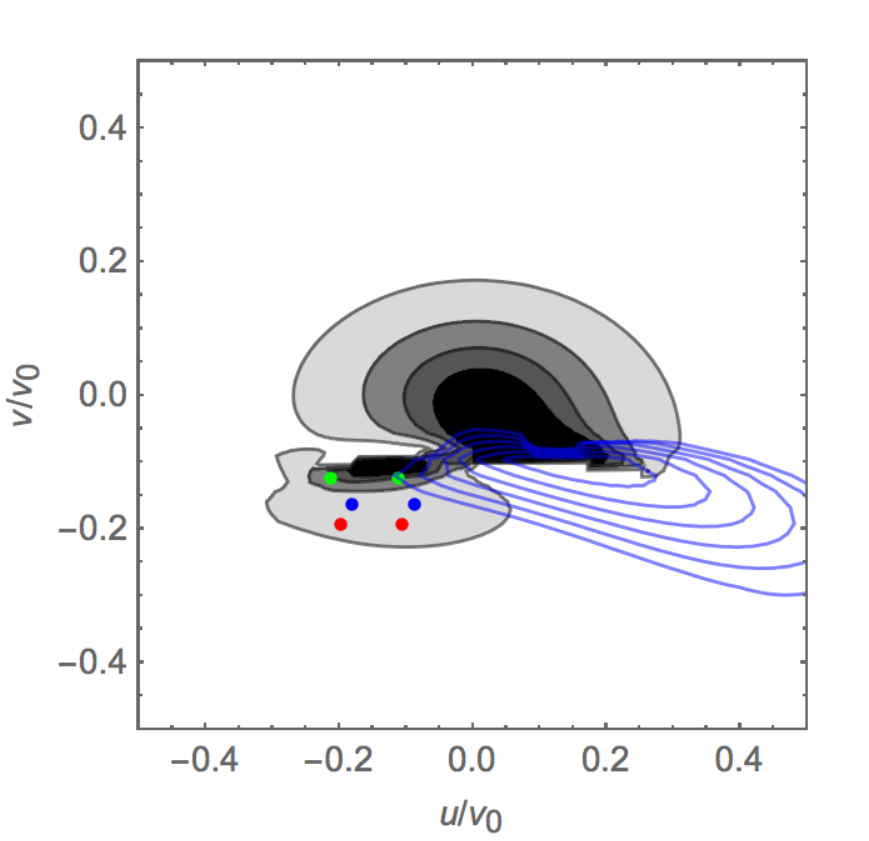}
  \caption{As in the central panel of \Fig{fig:vRvphi_ang}, but this
    time including the normalized energy of the pendulum
    $\eta=\Ep/\Vmax$, for orbits trapped at the bar's OLR. In
    particular, the blue contours correspond to
    $\eta_i=\etamin+i\Delta\eta$, for $i=1,...,6$, $\etamin$ the
    minimum of $\eta$, and $\Delta\eta=(1-\eta)/10$.}
  \label{fig:Ep}
\end{figure}
Then, in \Fig{fig:Ep} we reproduce the central plot
($\phib=25\degree$) of \Fig{fig:vRvphi_ang}, and we superpose on top
of it the contours of pendulum energy $\Ep$ for the OLR (blue
contours). We see that the contours representing the strongest
trapping coincide with the `horn' region that we mentioned in
\Sec{sect:bim}. Indeed, it seems that the overdensity that several
authors found in their simulations with the same fast bar models
\citep[e.g.,][]{Dehnen2000} could be explained by the orbits trapped to
the OLR resonance. \cite{MonariPhD} came to the same conclusion, using
however a different method, the numerical Fourier analysis of orbits
in the simulations.

\begin{figure}
  \centering 
  \includegraphics[width=0.49\columnwidth]{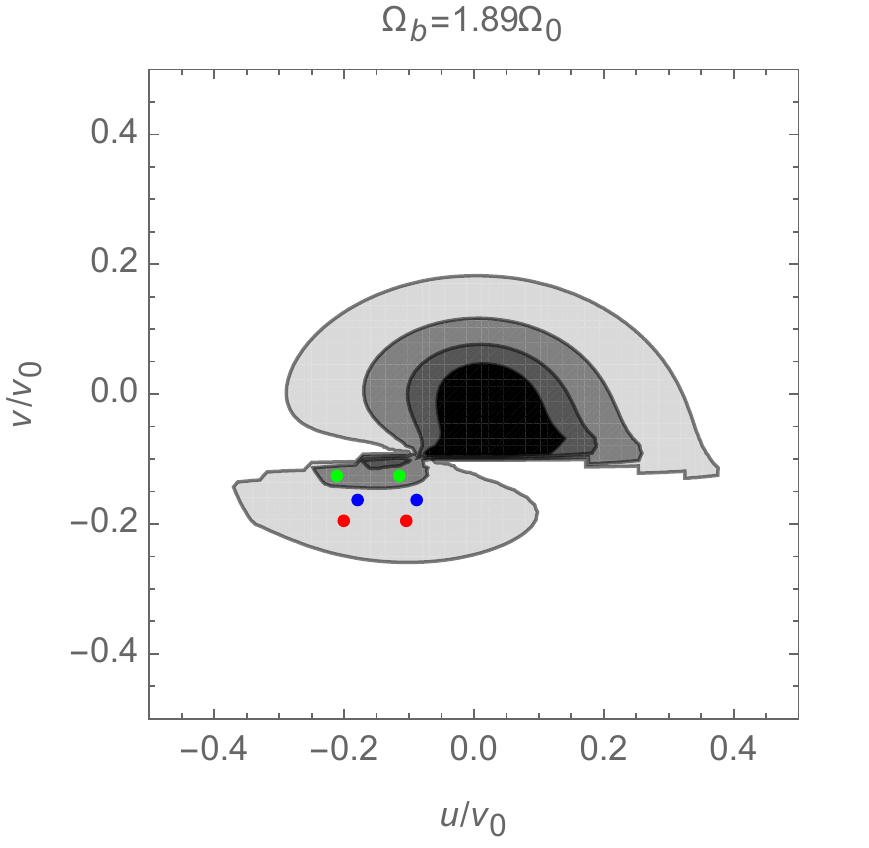}
  \includegraphics[width=0.49\columnwidth]{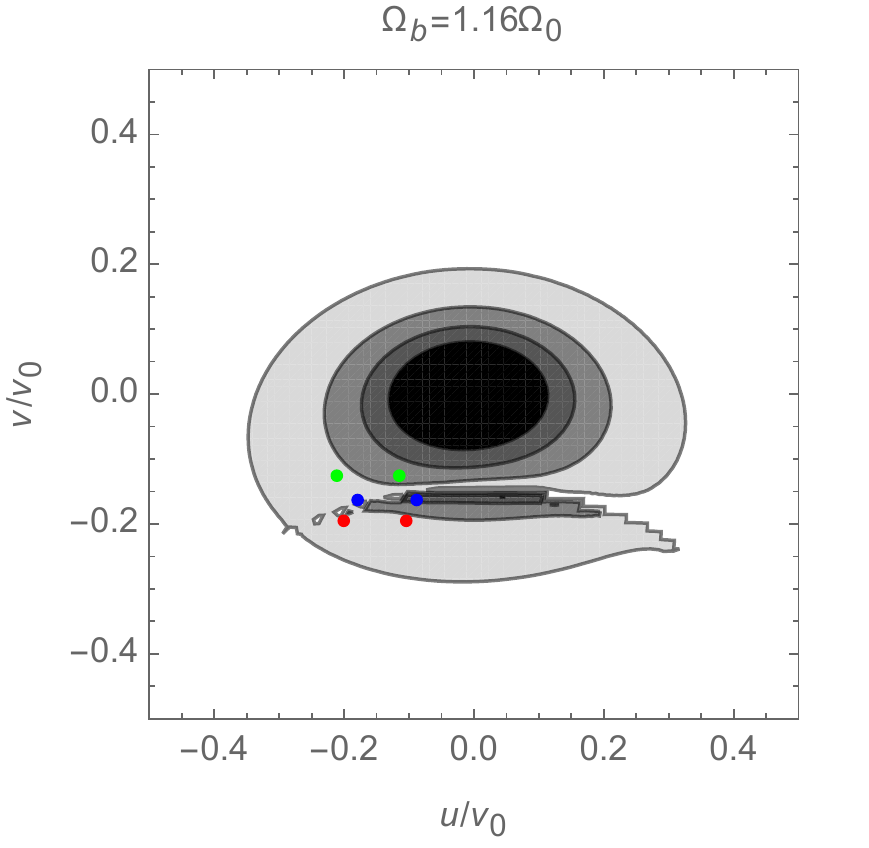}
  \caption{Velocity distribution to the first order (i.e. $f = f_0 +
    f_1$) in the $(u,v)$ plane at $(R,\phi,z) = (R_0, 0, 0)$ for a
    fast ($\Omb=1.89\Omega_0$) and slow bar model
    ($\Omb=1.16\Omega_0$), $\phib=25\degr$, and $\alphab=0.01$, where
    $f_0$ and $f_1$ are given in
    \Eqs{eq:eDF}{eq:eDF1}.}\label{fig:eDF}
\end{figure}
Finally, we note that we assumed throughout this paper a rather
  simple form for the background axisymmetric DF $f_0$. We thus also
  show the results (\Fig{fig:eDF}) for two different values of the
  pattern speed for a more involved axisymmetric DF, i.e.,
  \begin{equation}\label{eq:eDF}
    f_0=\sum_i w_if_{0,i}.
  \end{equation}
  Here, $f_{0,i}$ are quasi-isothermal DFs of the same kind of those
  used in M16 and in the rest of this work, but each of them has a
  different velocity dispersion $\tilde{\sigma}_{R,i}(R_0)$ (see
  equation~42 of M16). Each $f_{0,i}$ represents a stellar population
  with a different age $\tau_i$, which is related to the velocity
  dispersion via
  \begin{equation}
    \tilde{\sigma}_R(R_0)=\sigma_0\pare{\frac{\tau_i+\tau_1}{\taum+\tau_1}}^\beta,
  \end{equation}
  where $\sigma_0=40.1\kmsec$, $\tau_1=0.01\Gyr$, $\taum=10\Gyr$, and
  $\beta=0.33$ \citep{Binney2012}. This means that a stellar
  population born now has a dispersion $\tilde{\sigma}_R(R_0)\approx 4
  \kmsec$ , whilst a 1 Gyr old population has a dispersion
  $\tilde{\sigma}_R(R_0)\approx 19 \kmsec$, and a 10 Gyr old
  population a dispersion $\tilde{\sigma}_R(R_0)=40.1 \kmsec$. The
  weights $w_i$ are also inspired by equation~12 of \cite{Binney2012},
  and are related to $\tau_i$ by
  \begin{equation}\label{eq:w}
    w_i=\frac{\Delta\tau_i \eexp^{\gamma\tau_i}}{\sum_i
      \Delta\tau_i \eexp^{\gamma\tau_i}},
  \end{equation}
where $\Delta\tau_i$ are age intervals in the range $[0,\taum]$ for
thin disc stars taken from Table~3 of \cite{Besancon}; $\tau_i$ is
the central age value of each bin, and $\gamma=0.117$
\citep[see][]{AumerBinney2009}. 

The total linear response $f_1$ to the bar perturbation is simply the
weighted sum of the linear responses of each sub-population $f_{1,i}$,
i.e.,
  \begin{equation}\label{eq:eDF1}
    f_1=\sum_i w_if_{1,i}.
  \end{equation}

\Fig{fig:eDF} shows clearly that the results of our analysis in the
rest of this work, which uses a single population of velocity
dispersion $\tilde{\sigma}_R(R_0)=35\kmsec$ to study the effects of
the bar's pattern speed, still hold in the case of a more complex DF,
representing a reasonable superposition of stellar populations of
different ages and velocity dispersions.

\section{Conclusions}\label{sect:concl}

We presented the first application of the formalism developed in M16
to calculate, through perturbation theory, the effects of a
non-axisymmetric gravitational disturbance on an initially
axisymmetric DF, $f_0(\bJ)$, describing the phase-space density of
stars in a collisionless stellar system (i.e., governed by the
collisionless Boltzmann equation).

We extended the M16 formalism to second order (Section~3), and
concentrated on the effects of the Galactic bar on the DF in the Solar
neighbourhood. We checked whether a slow bar with pattern speed
$\Omega_0 \lesssim \Omb \lesssim 1.45 \Omega_0$ could reproduce the
observed bimodality of local velocity space. We concluded that no
feature in our modelled local velocity space could account for the
observed bimodality (Section~4). We checked whether second order
effects, or the additional effects of spiral arms, could help, and did
not find any configuration reproducing the bimodality. A fast bar with
$\Omb \approx 1.9 \, \Omega_0$, on the other hand, explains it nicely
(Section~5). \cite{BlandHGerhard} fixed their final value of the bar's
pattern speed at $\Omb= (1.48\pm0.31)\Omega_0$. Here we estimated
that, if Hercules is created by the bar's OLR, the pattern speed of
the bar cannot be less than $\Omb\approx1.8\Omega_0$ to be compatible
with the measured density peaks of the Hercules moving group.

In Section~3.3.3 and in all our figures, we also identified the
regions of resonant trapping in phase-space. This trapping should
affect the actual density of stars in the trapped zone compared to the
analytical models presented here, but does not strongly affect the
general distortion of phase-space itself, as numerical particle test
simulations with adiabatic growth of the bar
\citep[e.g.,][]{Monari2014} give the same result as our fast bar
models for the {\it shape} of the bimodality in local velocity space.
We note that, while such forward test-particle simulations can serve
as benchmarks to test analytical models like those presented here (see
M16), they do not allow to directly fit the data. Actually, the main
motivation of models based on analytical DFs is that they will indeed
allow to fit the data directly, with a few fitting parameters in the
perturbing potential as well as in the axisymmetric DF, by performing
a maximum-likelihood estimate of these parameters based on actual
kinematical data for a large set of individual stars. However, in
order to perform such a fully quantitative fit, our method will have
to be extended to better treat the DF for resonantly trapped
orbits. This will be the topic of a forthcoming paper.

Concerning the bimodality, let us also note that we did not try every
possible spiral arm configuration here, and cannot yet strictly
exclude that a similar structure as the locally observed bimodality
could be the result of spirals. Our results are generally in line with
the N-body simulations of \citet{Quillen2011} in which velocity
distributions created from regions just outside the bar's OLR more
closely resembled that seen in the solar neighbourhood. Nevertheless,
close inspection of the velocity distributions at other radii in these
simulations reveal spiral-related features which also slightly
resemble the Hercules stream, albeit at angles to the bar which do not
correspond to the present orientation of the bar in the Milky
Way. Also, \citet{Grand2014} showed that the outward radial migrators
behind their corotating spiral arms display lower-$v$ and negative-$u$
velocity (see their Fig.~4), hence providing a possible explanation
which will have to be inspected closely in the future. In any case,
the future DR2 and DR3 data releases from Gaia \citep{Gaia} should
allow a detailed investigation of phase-space structure outside of the
Solar neighbourhood, at different Galactic radii and azimuths, and
test our present conclusions about the pattern speed of the bar, since
any possible spiral-related features in velocity space would not
follow the same evolution at different radii and azimuths. Such a test
might actually already be possible by combining the Gaia DR1 with
existing spectroscopic surveys. We also note that the metallicity
patterns in local stellar velocity space seem to also support our fast
bar models (Antoja et al. 2016, in preparation).

The three--dimensional density of red clump giants in the inner Galaxy
nevertheless clearly indicate the existence of a long, flat structure,
oriented at an angle of $\phib \sim 27\degree$ from the Galactic
centre-Sun direction and reaching out to a radius $\sim 5$~kpc. The
most natural explanation would be that this structure is not a long
bar but rather a loosely wound spiral coupled to the end of the
bar. If it has a pattern speed only somewhat smaller than the central
bar, it could be a good candidate to explain the observed double-peak
aspect of the Hercules stream, which is not reproduced even in our
fast bar models. On the other hand, it is known that small nuclear
bars with faster pattern speed than the main bar can be long-lived in
numerical simulations including a gaseous component, even without
resonance overlaps or mode coupling, if star formation remains
moderately active in the region of the nuclear bar
\citep[e.g.,][]{Wozniak2015}. However, we are not aware of any
simulation reproducing a stable long bar with lower pattern speed than
its central counterpart and similar in size to the structure observed
in the inner 5~kpc of the Milky Way (hence about twice the disc
scale-length). We would thus {\it a priori} favour a loosely wound
spiral structure to explain the photometric observations.

\section*{Acknowledgements}
We thank the anonymous referee for the constructive reports. This work
has been supported by a postdoctoral grant from the {\it Centre
  National d'Etudes Spatiales} (CNES) for GM.

\bibliographystyle{mn2e}
\bibliography{mn-jour,f1_barbib}

\begin{appendix}

\end{appendix}

\label{lastpage}

\end{document}